\documentclass[twocolumn, balance]{aastex631}

\usepackage{graphicx}
\usepackage{amssymb}
\usepackage{natbib}
\usepackage{xspace}
\usepackage{appendix}

\newcommand{\Msun}{M$_\odot$\xspace}

\newcommand{\mosfit}{{\tt MOSFiT}\xspace}
\newcommand{\rubinsim}{{\tt rubin-sim}\xspace}

\begin{document}

\title{Prospects for Measuring Black Hole Masses using TDEs with the Vera C. Rubin Observatory}

\author[0000-0002-4235-7337]{K. Decker French}
\altaffiliation{Corresponding Author}
\affiliation{Department of Astronomy, University of Illinois, 1002 W. Green St., Urbana, IL 61801, USA} 
\affiliation{Center for Astrophysical Surveys, National Center for Supercomputing Applications, Urbana, IL 61801, USA}
\email{deckerkf@illinois.edu}

\author[0000-0001-6350-8168]{Brenna~Mockler}
\affiliation{The Observatories of the Carnegie Institute for Science, 813 Santa Barbara St., Pasadena, CA 91101, USA}

\author[0000-0003-1714-7415]{Nicholas Earl}
\affiliation{Department of Astronomy, University of Illinois, 1002 W. Green St., Urbana, IL 61801, USA} 
\affiliation{Center for Astrophysical Surveys, National Center for Supercomputing Applications, Urbana, IL 61801, USA}

\author{Tanner Murphey}
\affiliation{Department of Astronomy, University of Illinois, 1002 W. Green St., Urbana, IL 61801, USA} 
\affiliation{Center for Astrophysical Surveys, National Center for Supercomputing Applications, Urbana, IL 61801, USA}

\begin{abstract}

Tidal Disruption Events (TDEs) provide an opportunity to study supermassive black holes that are otherwise quiescent. The Vera C. Rubin Legacy Survey of Space and Time will be capable of discovering thousands of TDEs each year, allowing for a dramatic increase in the number of discovered TDEs. The optical light curves from TDEs can be used to model the physical parameters of the black hole and disrupted star, but the sampling and photometric uncertainty of the real data will couple with model degeneracies to limit our ability to recover these parameters. In this work, we aim to model the impact of the Rubin survey strategy on simulated TDE light curves to quantify the typical errors in the recovered parameters. Black hole masses $5.5< \log M_{\rm BH}/M_\odot < 8.2$ can be recovered with typical errors of 0.26 dex, with early coverage removing large outliers. Recovery of the mass of the disrupted star is difficult, limited by the degeneracy with the accretion efficiency. Only 57\% of the cases have accurate recovery of whether the events are full or partial, so we caution the use this method to assess whether TDEs are partially or fully disrupted systems. Black hole mass measurements obtained from Rubin observations of TDEs will provide powerful constraints on the black hole mass function, black hole -- galaxy co-evolution, and the population of black hole spins, though continued work to understand the origin of TDE observables and how the TDE rate varies among galaxies will be necessarily to fully utilize the upcoming rich data set from Rubin. 

\end{abstract}


\section{Introduction} \label{sec:intro}

Tidal Disruption Events (TDEs) can be observed when stars pass too close to supermassive black holes (SMBHs) and are partially or fully disrupted by tidal forces. TDEs will briefly illuminate otherwise-quiescent SMBHs as an accretion disk forms and dissipates, providing a powerfully complimentary probe of the SMBH population to the tests enabled by the long-lived accretion disks in Active Galactic Nuclei (AGN). 

Currently, more than a dozen TDEs are discovered each year, primarily through optical time-domain surveys. Starting this year, the Vera C. Rubin Observatory Legacy Survey of Space and Time will begin to revolutionize the field of time-domain observations, with predictions for thousands of TDEs discoverable each year \citep{vanVelzen2011, Bricman2020}. Rubin will survey the Southern sky in six bands ($ugrizy$) and produce $\sim$10 million alerts each night. 

The photometric information from Rubin will provide valuable constraints on the systems producing TDEs. Depending on the emission mechanisms acting in TDEs, one or several observables will be sensitive to the mass of the disrupting black hole. The UV/optical emission from TDEs has several proposed origins, with theories primarily attributing this emission to reprocessed emission from the accretion-disk scale \citep[e.g.,][]{Metzger2016, Dai2018} or shocks at apocenter from the formation of the accretion disk \citep[e.g.,][]{Piran2015}. Recent simulations have demonstrated that the emission we observe may be reprocessed through an optically thick wind launched by a combination of the accretion disk and shocks at smaller pericenter or disk scales \citep{Huang2023, Steinberg2024, Price2024}. If the observed optical light is driven by fallback-limited accretion (or fallback-limited shocks), the timescale of the event scales as the square root of the black hole mass \citep[hereafter M19]{Guillochon2013, Mockler2019}. This model has been used to infer black hole masses among other parameters for a wide range of TDEs \citep[M19, ][]{Nicholl2022}. Additionally, the peak luminosity (\citealt{Mummery2024}, or peak luminosity with temperature, \citealt{Ryu2020}), as well as the luminosity of disk emission in the plateau phase \citep{Mummery2024} have also been proposed as tracers of black hole masses using TDEs.

The proposed cadence strategies for Rubin are optimized for a wide range of science cases, from asteroid detection to supernovae. To facilitate these various science cases, the proposed cadence schedule will be intrinsically irregular in each band, with the aim to achieve a roughly 3-day cadence between any two bands. Most TDEs have relatively constant colors \citep{Holoien2016, vanvelzen2021}, which will make them relatively robust against uncertainties in this cadence, compared to other science cases with rapidly-changing colors such as supernovae. However, the cadence used by Rubin will nonetheless introduce uncertainties in the modeling of TDEs. In this work, we aim to model the light curves Rubin will measure for TDEs and determine how accurately their physical parameters can be inferred, given the expected sampling, depths, and photometric precision. In \S\ref{sec:methods} we describe the construction of our synthetic TDE light curves, and in \S\ref{sec:results} we present the recovered black hole masses, disrupted star masses, and impact parameters as inferred for our synthetic TDEs. In \S\ref{sec:discussion} we discuss the caveats and limitation of this work, compare to other Rubin predictions in the literature, and discuss the implications for using TDEs to study black hole -- galaxy co-evolution with Rubin. We conclude in \S\ref{sec:conclusions}.

\section{Synthetic TDE light curves}
\label{sec:methods}

\begin{figure}
\begin{center}
\includegraphics[width=0.5\textwidth]{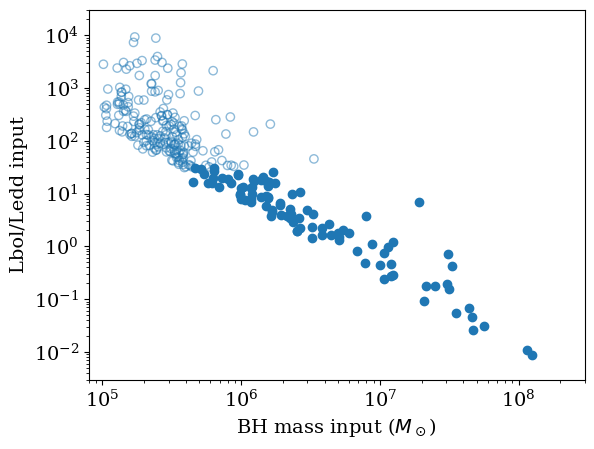}
\end{center}
\caption{Eddington fraction vs. black hole mass for the simulated TDEs we generate with \mosfit. We generate a sample of 300 synthetic TDEs, as described in the text. The majority of these have peak Eddington fractions $L_{\rm bol}/L_{\rm Edd} > 30$, which is higher than typically considered in TDE simulations and beyond what can be accurately modeled with \mosfit. We restrict our analysis going forward to only the 87 events with input $L_{\rm bol}/L_{\rm Edd} < 30$. The majority of simulated TDEs around black holes with $ \log M_{\rm BH}/M_\odot<5.5$ are highly super-Eddington and excluded from our sample. Because of this effect, we stress that our resulting population of simulated TDEs is no longer a complete model population covering the full range of likely black hole masses.
}
\label{fig:Lbolcut}
\end{figure}

\begin{figure*}
\begin{center}
\includegraphics[width=\textwidth]{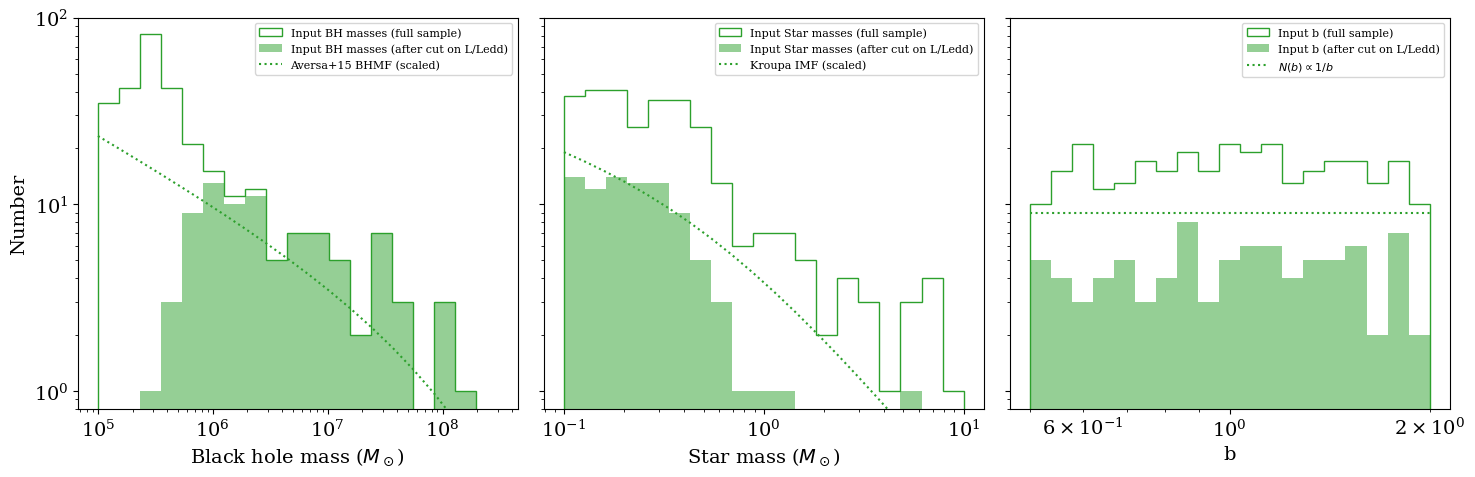}
\end{center}
\caption{Black hole masses (left), star masses (center), and scaled impact parameters ($b$) used to generate synthetic TDEs. The unfilled (filled) histogram shows the sample before (after) our cut on Eddington fraction. The dotted line in each case shows the scaled distributions from which we select each parameter. 
}
\label{fig:input}
\end{figure*}

\begin{table*}[]
    \centering
    \begin{tabular}{l c c}
    \hline
    Parameter &  Value/range & Sampling \\ 
    \hline
Black hole mass ($M_{\rm BH}$) &  $10^5-3\times10^8$ \Msun & \citet{Aversa2015} \\
Star mass ($M_\star$) & $0.1-10$ \Msun  & \citet{Kroupa2001}\\
Impact parameter ($b$) &   $0.5-2$ & $1/\beta$ distribution\\
Efficiency ($\epsilon$) &  0.1 & \\
Viscous delay time ($T_{\rm visc}$) & 0.001 days & \\
Photosphere power law constant  ($R_{\rm ph0}$) & 1 & \\
Photosphere power law exponent  ($l_{\rm photo}$) & 1 & \\
Host column density ($N_H$) & $10^{20}$ cm$^{-2}$ & \\
Variance & 0.1 & \\
Redshift ($z$) & $0-0.3$ & uniform sampling in volume \\
R.A., decl. & decl. $<30^\circ$ & uniform sampling in angular space \\
\hline
    \end{tabular}
    \caption{Description of free and fixed parameters used to model TDE lightcurves using \mosfit, as described in \S\ref{sec:methods}.}
    \label{tab:params}
\end{table*}

\begin{figure*}
\begin{center}
\includegraphics[width=0.49 \textwidth]{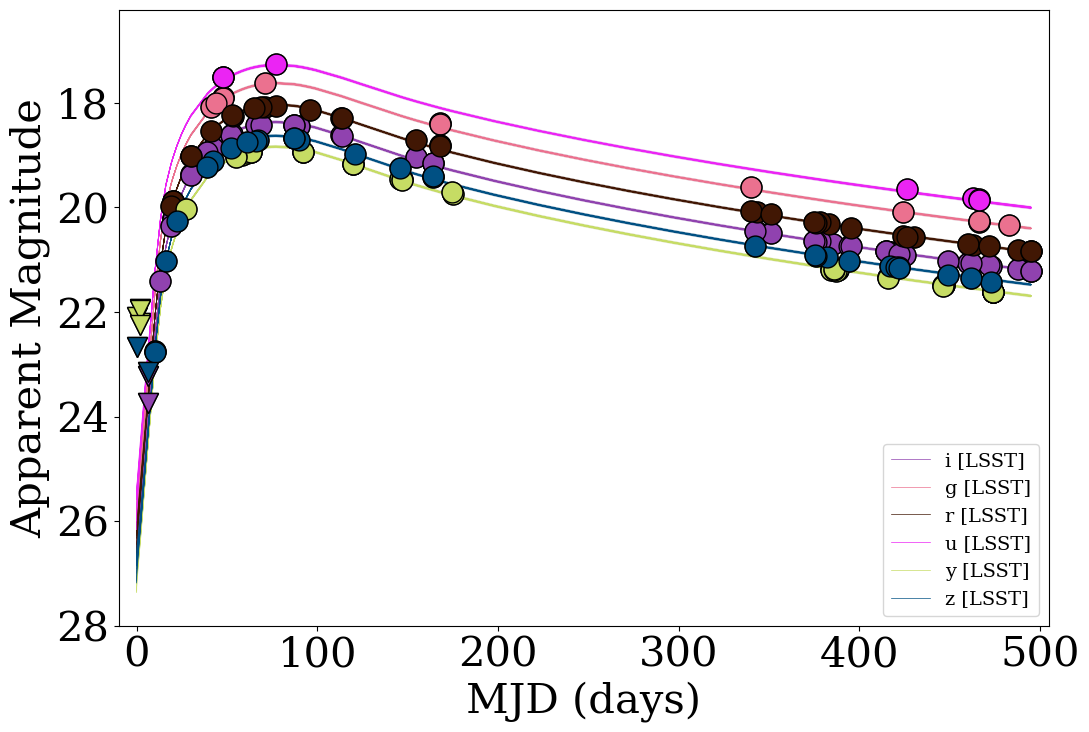}
\includegraphics[width=0.49 \textwidth]{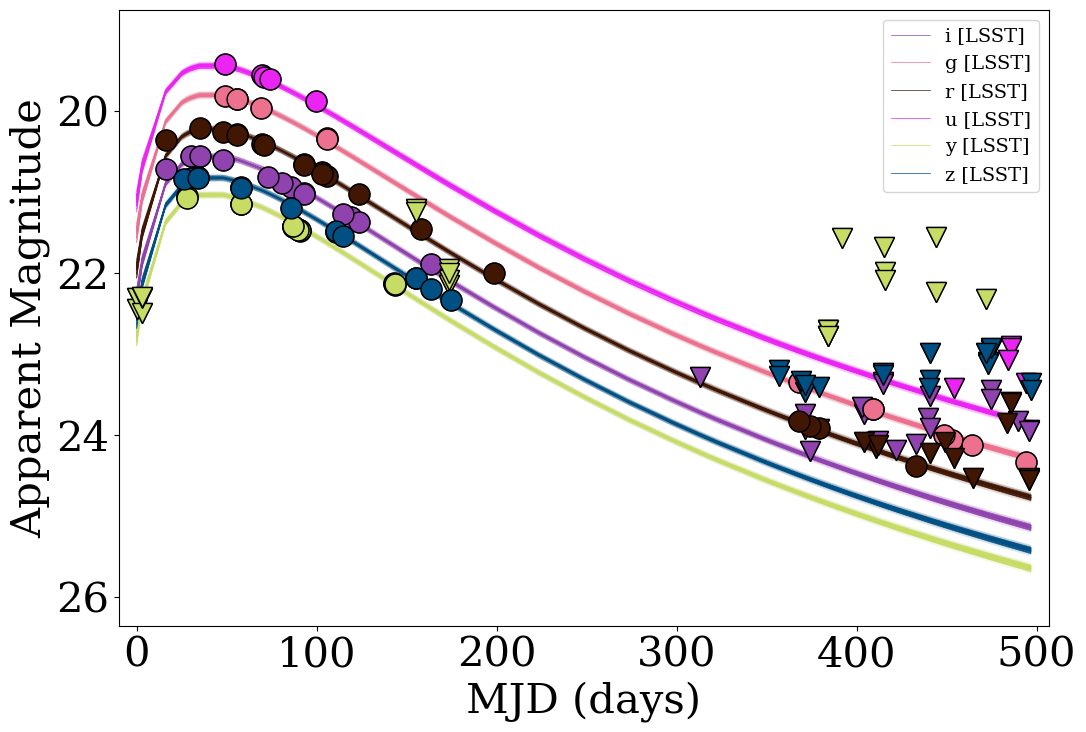}
\includegraphics[width=0.49 \textwidth]{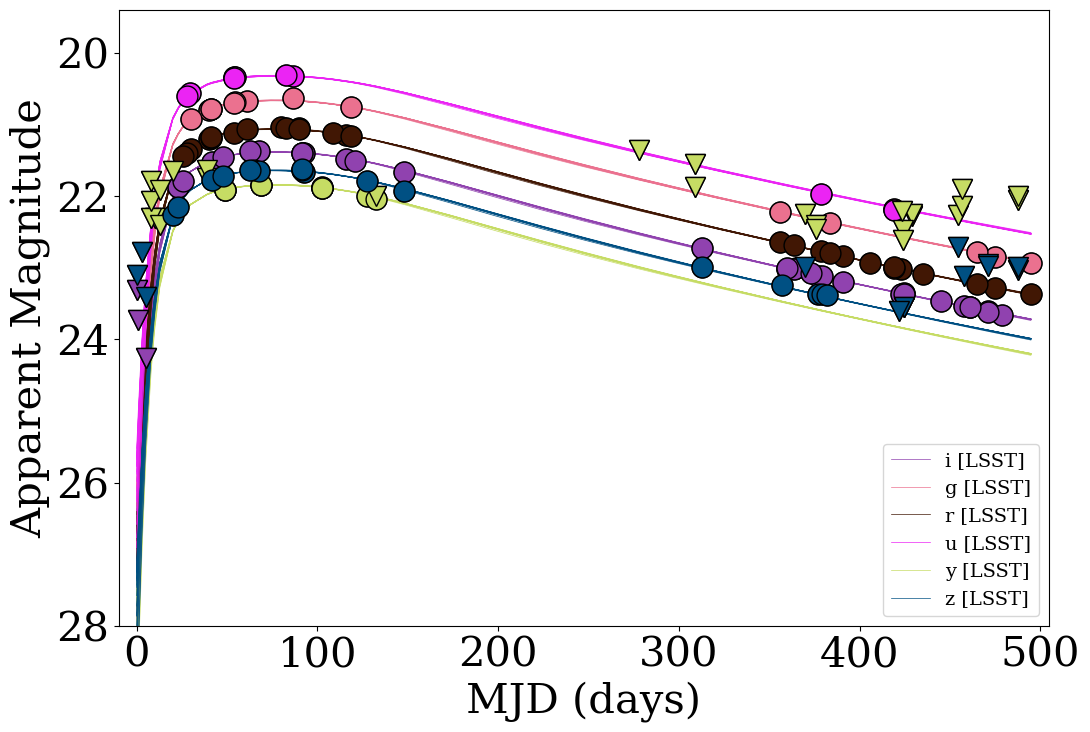}
\includegraphics[width=0.49 \textwidth]{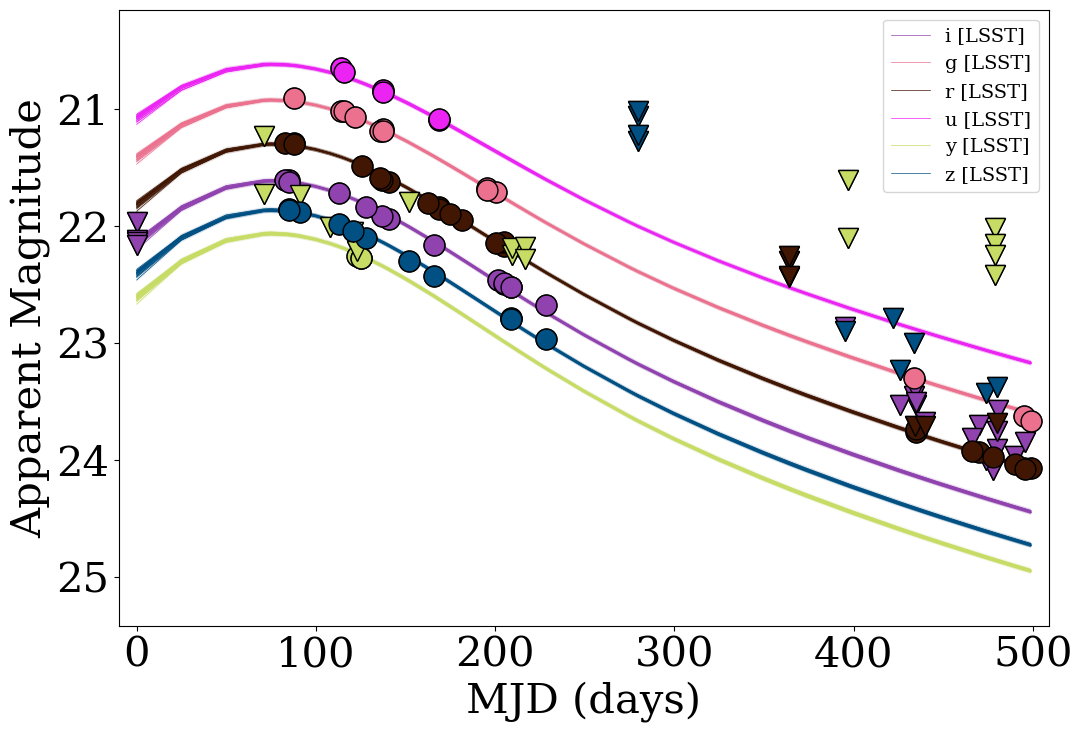}
\end{center}
\caption{Four example synthetic TDE light curves, constructed as described in \S\ref{sec:methods}. The detections, limits, seasonal gaps, and cadence reflect realistic observing conditions that will limit our ability to recover TDE parameters. Best-fit \mosfit results are shown as solid lines, obtained as discussed in \S\ref{sec:results}.}
\label{fig:lcexamples}
\end{figure*}

M19 have developed a model for predicting TDE light curves assuming the optical/UV light is reprocessed emission originating from fallback-limited accretion or shocks. Given a range of stellar structures, M19 use a suite of hydrodynamical simulations to determine the distribution of debris mass binding energy after a star is disrupted. The luminosity is assumed to follow the rate of mass fallback with some efficiency $\epsilon$.  The luminosity as a function of time will thus depend on the black hole mass, the impact parameter, the star mass, and the efficiency. M19 also allows for a viscous delay and models the resulting light emitted as a photosphere. The photosphere is modeled as a blackbody that evolves as the radius of the photosphere changes, with parameters for the radius normalization and power law exponent. The resulting model thus contains seven free parameters (in addition to a start time), as listed in Table \ref{tab:params}.

We use the implementation by M19 of their model within the Modular Open Source Fitter for Transients (\mosfit) to generate model TDE light curves from a range of parameters. We list the parameters, ranges or fixed values, and sampling distributions in Table \ref{tab:params}. We begin by using inverse transform sampling to randomly select black hole masses from an \citet{Aversa2015} black hole mass function, star masses from a \citet{Kroupa2001} initial mass function (IMF), and scaled impact parameters ($b$) from a $1/b$ distribution. Distributions of these parameters (in log space) are shown in Figure \ref{fig:input}. For the black hole masses and impact parameters, our sampling distributions differ from the priors used by \mosfit, which assumes a log prior on the black hole mass and a flat prior on the scaled impact parameter. We fix the remaining \mosfit parameters based on typical values from the sample of TDEs fit by M19 (Table \ref{tab:params}). 

This procedure results in a large number of model TDEs with luminosities much greater than can be effectively modeled using \mosfit. We generate 300 model TDEs, finding that only 87/300 have peak Eddington luminosity ratios $L_{\rm bol}/L_{\rm Edd} < 30$. Simulations of TDEs rarely include accretion rates this high, e.g., \citet{Thomsen2022} consider accretion rates of $7-24 \dot{M}_{\rm Edd}$. \mosfit imposes a soft boundary at $L_{\rm Edd}$, and thus highly super-Eddington events will not be well-recovered. Simulated light curves of these super-Eddington events rise quickly and plateau at high luminosities, which cannot be well-fit by \mosfit (the recovered black hole masses are not well-correlated with the input masses) and are dissimilar to the observed population of TDEs. We restrict our analysis going forward to only the 87 events with input $L_{\rm bol}/L_{\rm Edd} < 30$. In Figure \ref{fig:Lbolcut} we demonstrate the effect that this cut has on our black hole mass input distribution. The majority of simulated TDEs around black holes with $ \log M_{\rm BH}/M_\odot<5.5$ are highly super-Eddington and excluded from our sample. This cut also preferentially removes TDEs from the highest disrupted star masses. Because of this effect, we stress that our resulting population of simulated TDEs is no longer a complete model population covering the full range of possible black hole masses.

Next, we use the {\tt rubin-sim} \citep{Ivezic2019, Connolly2010, Connolly2014, Delgado2016} baseline simulations (v3.4) to determine the cadence, filters, and depths for our synthetic observations. We select coordinates sampled evenly in angular space, excluding coordinates too far north for Rubin (requiring dec $<30^\circ$). For each coordinate, we use {\tt rubin-sim} to determine the simulated observation times, $5\sigma$ depths, and filters. 

Each synthetic TDE is assigned a distance, using a redshift range $0<z<0.3$, selected to be uniform in volume. This corresponds to a luminosity distance $D_L \lesssim 1500$ Mpc. While Rubin will be capable of discovering TDEs out to larger distances, these closer TDEs will represent our best-studied sample for which multi-wavelength followup and spectroscopic classification will be feasible for moderate sample sizes. For example, TiDES (Time Domain Extragalactic Survey) will use 4MOST (4-metre Multi-Object
Spectroscopic Telescope) to classify transients brighter than 22.5 mag, which will enable several hundred TDE classifications, primarily at $z<0.3$ \citep{Frohmaier2025}.

Using the \mosfit model and \rubinsim information for each synthetic TDE, we generate a synthetic lightcurve by associating each model TDE to a single randomly-chosen distance and position. We consider TDEs that begin during the first month of the survey, which experience a range of seasonal gaps due to their sky positions. For each observation, if the modeled magnitude is fainter than the $5\sigma$ depth, we consider the observation a non-detection and use the $5\sigma$ depth as an upper limit instead. If the source is brighter, we add a random number using a Gaussian centered at 0 with width 0.01 mag to add random noise at the level of the anticipated Rubin photometric accuracy\footnote{The band-to-band photometric calibration including the $u$ band has a design specification of 10 mmag and a stretch goal of 5 mmag (LSST Science Requirements Document, LSST Document LPM-17 \url{https://ls.st/LPM-17}).}. Examples of the synthetic TDE light curves are shown in Figure \ref{fig:lcexamples}.

\section{Recovery of Black hole mass, stellar mass, and scaled impact parameter}
\label{sec:results}

\begin{figure*}
\begin{center}
\includegraphics[width=0.49 \textwidth]{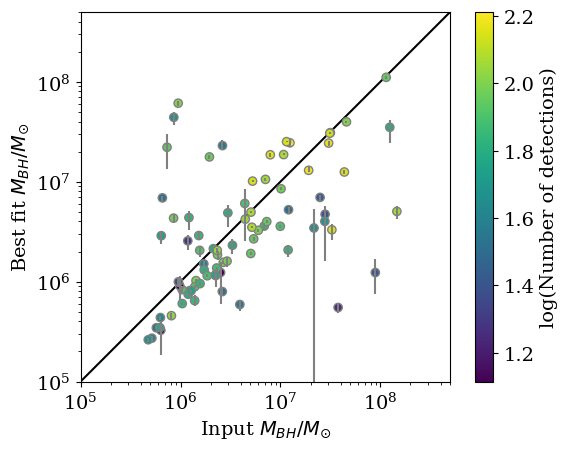}
\includegraphics[width=0.49 \textwidth]{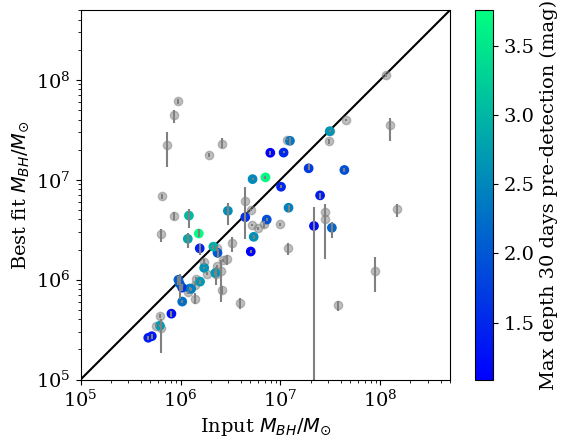}
\end{center}
\caption{Recovery of black hole mass by \mosfit fitting of synthetic TDE observations, colored by the number of detections (left) and maximum depth achieved (in any band) in the 30 days prior to the first detection (right). Grey circles are cases with no limits $>1$ mag below the first detection in the 30 days prior to that first detection. The black hole masses have a median difference of 0.26 dex between the input and recovered values. The total number of detections primarily scales with the black hole mass, as the higher black hole mass events are brighter and longer-lived. The synthetic TDEs with good early coverage are the most accurately recovered. If we consider only synthetic TDEs with limits of $>1$ mag below the first detection in the 30 days prior, the black hole mass recovery is accurate to 0.25 dex and we remove cases with extreme $>1$ dex differences from the input black hole mass. }
\label{fig:bhmass}
\end{figure*}

\begin{figure*}
\begin{center}
\includegraphics[width=0.49 \textwidth]{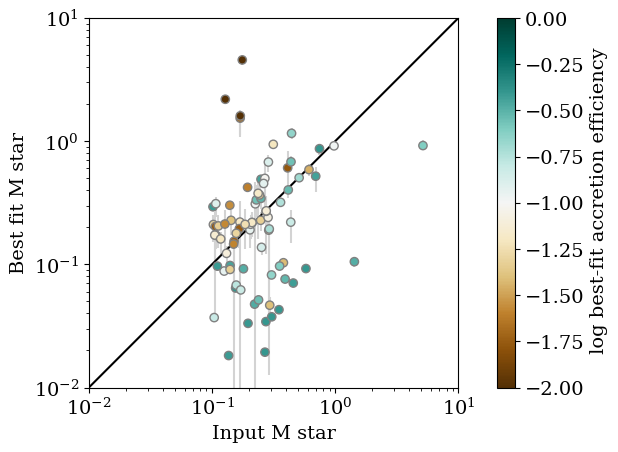}
\includegraphics[width=0.49 \textwidth]{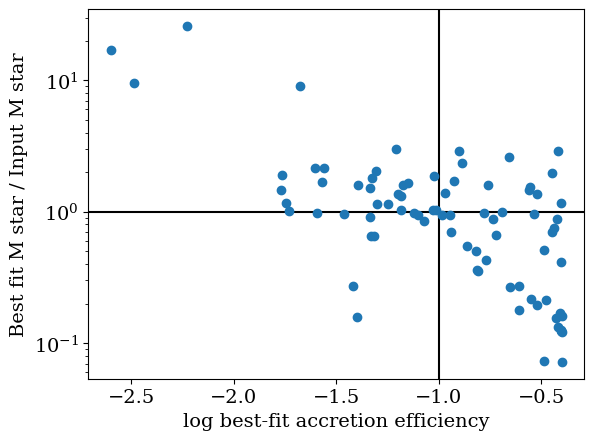}
\end{center}
\caption{Left: recovery of disrupted star mass by \mosfit fitting of synthetic TDE observations, colored by the best-fit accretion efficiency. Right: ratio of recovered to input star mass vs. accretion efficiency. The mass of the disrupted star is highly degenerate with the efficiency of luminosity produced by mass accretion $\epsilon$ \citep{Mockler2021}. Because we fix the efficiency of the synthetic TDEs to be $\epsilon = 0.1$, while the efficiency is a free parameter in our \mosfit recovery fits, the resulting best-fit star masses are degenerate with the efficiency. Cases where the recovered star mass is higher than the input are typically those with lower recovered efficiencies $\epsilon < 0.1$, and vice versa. }
\label{fig:starmass}
\end{figure*}

\begin{figure}
\begin{center}
\includegraphics[width=0.49 \textwidth]{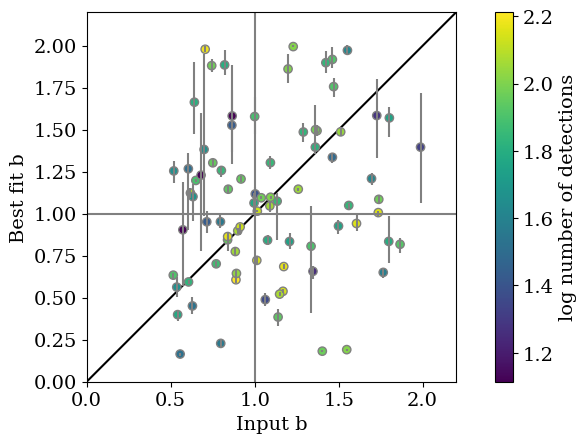}
\end{center}
\caption{Recovery of scaled impact parameter $b$ from \mosfit fitting of synthetic TDE observations, colored by the number of early detections. $b=1$ corresponds to the threshold for a full disruption, with $b<1$ indicating a partial disruption. 57\% of the cases have accurate recovery of whether the events are full or partial, so we caution the use of best-fit $b$ measurements in Rubin observations to assess whether TDEs are partially or fully disrupted systems.  }
\label{fig:beta}
\end{figure}

After constructing the synthetic light curves, we use \mosfit to fit the synthetic TDEs, to compare the recovered parameters with our input parameters. We use dynesty \citep{Speagle2020} as our sampler, and consider only the first 500 days of data. As mentioned above, we only consider the 87 cases with peak Eddington fractions $L_{\rm bol}/L_{\rm Edd} < 30$. Of these 87, 81/87 have at least 10 detections, so we exclude the six poorly-sampled events. 

We plot the input vs. recovered black hole masses in Figure \ref{fig:bhmass}. The black hole masses have a median difference of 0.26 dex between the input and recovered values $\rm med (|\log M_{BH, Output} - \log M_{BH, input}|)$. This is comparable to the 0.2 dex systematic error estimated by \citet{Mockler2019}, before introducing the observational uncertainties we consider here.  When considering the distribution of these systematic differences, a \citet{Dagostino1971,Dagostino1973} test for normality shows that the distribution of systematic differences is not consistent with a Gaussian distribution. We investigate the trends in accuracy and data quality. The total number of detections primarily scales with the black hole mass, as the higher black hole mass events are brighter and longer-lived. The synthetic TDEs with good early coverage of the shape and timescale of the rise are the most accurately recovered. We parameterize the useful early coverage by measuring the maximum depth limit measured in the 30 days before the first detection. Cases with no useful limits before detection are typically the worst-recovered. If we consider only synthetic TDEs with limits of $>1$ mag below the first detection in the 30 days prior, the median black hole mass recovery is accurate to 0.25 dex, we remove cases with extreme $>1$ dex differences between the input and best-fit black hole mass, and the distribution of systematic differences is indistinguishable from a Gaussian distribution.

The mass of the disrupted star is highly degenerate with the efficiency of luminosity produced by mass accretion $\epsilon$ \citep{Mockler2021}. Because we fix the efficiency of the synthetic TDEs to be $\epsilon = 0.1$, while the efficiency is a free parameter in our \mosfit recovery fits, the resulting best-fit star masses are degenerate with the efficiency. We compare the input and recovered star masses in Figure \ref{fig:starmass}, colored by the best-fit efficiency. Cases where the recovered star mass is higher than the input are typically those with lower recovered efficiencies $\epsilon < 0.1$, and vice versa. The median difference between the input and recovered star mass is 0.24 dex. Given the fixed input efficiency and the degeneracy noted above, we caution that the true star mass systematic errors could be higher. We do not see clear trends in the accuracy of star mass recovery with the number of detections or max depth pre-discovery. 

We also consider the recovery of the scaled impact parameter $b$. As defined in M19, $b=1$ corresponds to the threshold for a full disruption regardless of the polytrope used for the stellar structure. Cases where $0<b<1$ will be partial disruptions. Given the ratio of the tidal radius $R_t$ to radius at pericenter $R_p$, $\beta \equiv R_t/R_p$, $b=2$ corresponds to $\beta=2.5$ for a star with polytrope index $\gamma=5/3$ (typical of stars with mass $< 0.3$ \Msun or $>20$ \Msun), and to $\beta=4$ for a star with polytrope index $\gamma=4/3$ (typical of intermediate mass stars with mass $\sim 1-15$ \Msun). The input vs. recovered values of $b$ are shown in Figure \ref{fig:beta}. The typical difference in recovered vs. input $b$ is 0.42 (or, 0.15 dex). We do not see a significant improvement in the value of $b$ recovered when restricting the data quality using the number of data points, or the number or depth of early data points. The difficulty in recovering $b$ may be connected to the difficulty in accurately constraining the power-law decay slope of the bolometric luminosity after the peak of the flare, particularly in the absence of UV data to constrain the peak of the optical/UV blackbody. Varying late-time slopes have been reported for similar data on the same events \citep{vanvelzen2021, Hammerstein2023}, demonstrating the difficulty of the measurement using optical photometry. Our results here show that similar difficulties in inferring $b$ are likely to occur when using Rubin data, without additional UV observations. 
57\% of the cases have accurate recovery of whether the events are full or partial, so we caution the use of best-fit $b$ measurements in assessing whether TDEs are full or partial. Indeed, many of the events thought to be full TDEs may be partial, as demonstrated by the repeat TDE in AT2022dbl \citep{Makrygianni2025}.

\begin{figure*}
\begin{center}
\includegraphics[width=0.8 \textwidth]{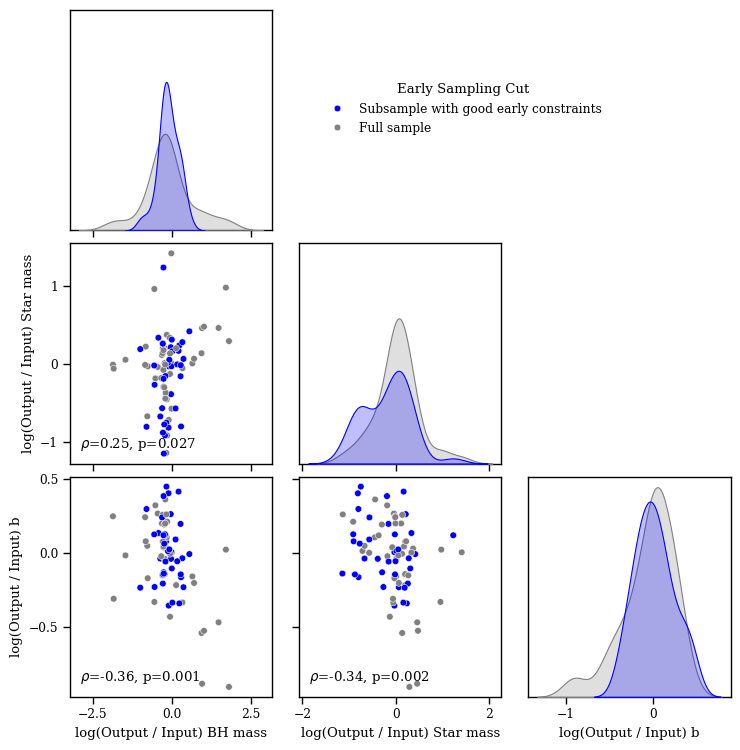}
\end{center}
\caption{Correlations between the bias in recovered vs. input black hole mass, star mass, and scaled impact parameter. Simulated TDEs with a lower best-fit $b$ than simulated tend to have recovered black hole masses and star masses biased high. Simulated TDEs with higher black hole mass recovered tend to also have a higher recovered star mass. Grey points represent the full sample of TDEs as shown in Figures \ref{fig:bhmass}-\ref{fig:beta} and blue points represent only the TDEs for which we have good early sampling, defined as having limits $>1$ mag below the detection in the 30 days prior. Normalized histograms are shown for each parameter. For the black hole masses, good early sampling helps to remove the large outliers and make the resulting distribution more Gaussian.}
\label{fig:combo}
\end{figure*}

The recovery of these parameters is correlated, as we demonstrate in Figure \ref{fig:combo}. We compare the difference in recovered log black hole mass, log star mass, and log scaled impact parameter. Simulated TDEs with a lower best-fit $b$ than simulated tend to have recovered black hole masses and star masses biased high relative to the input. Simulated TDEs with higher black hole mass recovered tend to also have a higher recovered star mass.

\section{Discussion}
\label{sec:discussion}

\subsection{Caveats and limitations}

The sample of synthetic TDEs we consider is not representative of the true observed population of TDEs for several reasons. The first is the issue we discuss in \S\ref{sec:methods}, where a large fraction of the TDEs simulated using a realistic IMF and black hole mass function are highly super-Eddington, and thus cannot be well-modeled by fallback-limited accretion. Super-Eddington systems may form fundamentally different accretion flows \citep[e.g.,][]{Zhang2025}. 

Furthermore, we do not consider the variation in the TDE rate as a function of black hole mass. Theoretical work using the structure of the inner regions of galaxies over a wide range of galaxy mass predicts a TDE rate that varies with black hole mass, peaking at $M_{\rm BH} \sim 10^6$ \Msun \citep{Chang2025, Hannah2025}. \citet{Chang2025, Hannah2025} also consider the likely variation of TDE impact parameter with black hole mass, with high $b$ events more likely around lower mass black holes, which would impact the observed properties of a true population sample. 

\begin{figure*} 
\begin{center}
\includegraphics[width=0.49 \textwidth]{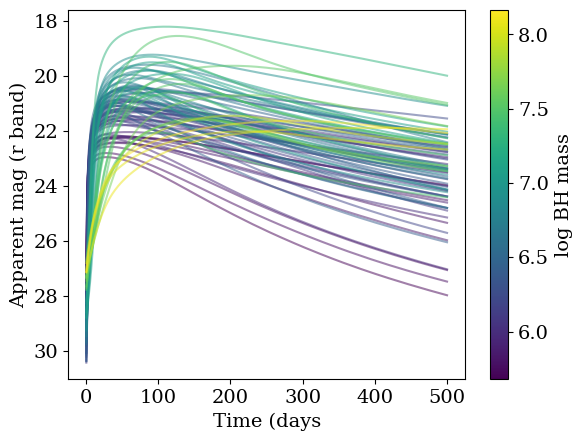}
\includegraphics[width=0.49 \textwidth]{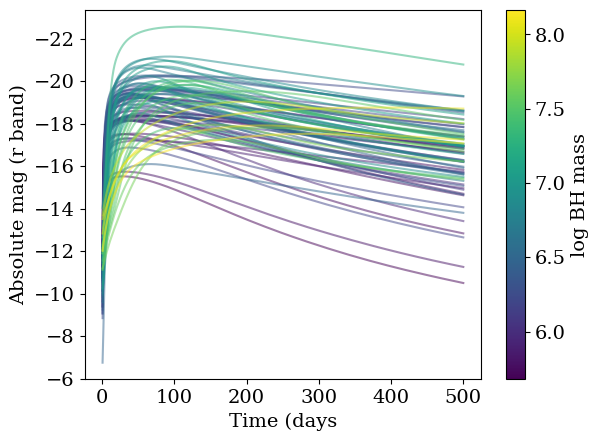}
\end{center}
\caption{Model light curves for the simulated TDEs (left: apparent magnitude; right: absolute magnitude), colored by the input black hole mass. Higher mass TDEs have longer durations and brighter peaks, although the relation between black hole mass and luminosity will have significant dependence on the accretion efficiency and star masses assumed. }
\label{fig:lc_colorbymass}
\end{figure*}

In this work, we consider the emission from fallback-limited accretion (or fallback-limited shocks) observed at early times from the flare portion of the TDE. However, other possibilities for measuring black hole masses using other observable features have been proposed. \citet{Ryu2020} have proposed a method (TDEmass), that uses the peak luminosity and temperature of a TDE to infer the black hole mass, assuming the emission is generated by shocks at the apocenter of the forming accretion disk. \citet{Sarin2024} have proposed a cooling envelope model that has also been used to constrain black hole masses through light curve modeling. \citet{Mummery2024} use late-time observations of TDEs, after the initial flare has faded, to measure the plateau luminosity from the viscously-spreading accretion disk, which provides an independent constraint on black hole mass (see \citealt{Ramsden2025} for a discussion of modeling black hole populations in Rubin using the plateau method). Furthermore, the easy-to-measure peak TDE luminosity has been shown to scale with the black hole mass. \citet{Mummery2024} provide a calibration of peak luminosity to black hole mass based on plateau luminosity measurements. In Figure \ref{fig:lc_colorbymass} we show the \mosfit-modeled light curves for our simulated TDEs in apparent and absolute $r$-band magnitude, colored by black hole mass. The more massive black holes typically have more luminous TDEs, but there is significant scatter. Similarly, \citet{Mummery2024} find 0.53 dex scatter in the relation between peak luminosity and black hole masses inferred from plateau measurements. Peak-luminosity-inferred black hole masses will be measurable for a larger number of TDEs in the era of Rubin, albeit with larger uncertainties than from modeling the TDE flare itself.

\subsection{Comparison to previous work}

Several recent works have modeled TDE observations with Rubin. \citet{Bricman2020} consider a population of simulated TDEs modeled using \mosfit, varying black hole mass and $b$, keeping the other free parameters (including star mass) constant. TDEs are populated within simulated galaxies, with black hole masses chosen based on a range of possible black hole mass functions. \citet{Bricman2020} evaluate the number of detections that will be possible for each TDE, finding a systematic effect with lower mass black holes having a decreased detection fraction. Our redshift cut limits the number of TDEs to primarily those that will be well-observed; 10-20\% of the detected TDEs from \citet{Bricman2020} lie within $z<0.2$. In addition to the uncertainty in the recovered black hole masses that we consider in this work, obtaining an accurate black hole mass function from TDE observations will be needed to incorporate the systematic effects described by \citet{Bricman2020}. 

\citet{Gomez2023} also simulates a population of TDEs for both Rubin and Roman, using the population of \mosfit-inferred parameters from the TDEs studied by \citet{Nicholl2022}. Using a uniform distribution in comoving space, well-sampled TDEs are predicted to occur out to $z\sim 2$ in the Rubin data. For higher redshift cases, the black hole mass recovery will likely become worse, with higher systematic uncertainties.

\subsection{Prospects for measuring black hole-galaxy co-evolution with Rubin}

\subsubsection{The black hole mass function}

A key goal of measuring supermassive black hole masses with TDEs will be to infer the underlying black hole mass function. Using TDEs to trace black holes is complimentary to analyses that use AGN (such as \citealt{Aversa2015}, used here to model our black hole mass function). AGN and TDEs will have different biases in tracing black holes, allowing us to test for systematic effects. Furthermore, pushing the black hole mass function into the regime of intermediate mass black holes will allow for tests of black hole seeding mechanisms \citep{Greene2020}. If black hole seeds are massive and rare, the black hole mass function will plateau and decline towards lower masses, while if black hole seeds are light and common, the black hole mass function will continue to rise down toward stellar-mass black holes. 

For black holes with masses $>10^6$ \Msun, the precision of our recovered black hole masses of $\sim0.26$ dex will allow for TDE rates as a function of black hole mass to be well constrained. However, it is clear from \S\ref{sec:methods} that TDEs in lower mass  $<10^6$ \Msun systems can be easily super-Eddington for the typical population of stars, especially as a higher fraction of the TDEs will be full disruptions \citep{Chang2025, Hannah2025}. It will be necessary to fully understand the observational signatures of TDEs around IMBHs in order to incorporate these systems into the black hole mass function. 

Furthermore, obtaining a black hole mass function from the observed TDE rate function will require detailed theoretical work \citep{Chang2025, Hannah2025} to understand how the intrinsic TDE rate varies with black hole mass. Some of the lower black hole mass TDEs studied thus far \citep[e.g. AT2020neh, ][]{Angus2022} may have enhanced TDE rates due to their centrally concentrated stellar distributions, with higher rates than expected for typical dwarf galaxies.

\subsubsection{Black hole growth during quenching}

Tracing the properties of host galaxies undergoing rapid evolution can be used to test how and when supermassive black holes grow. Black hole growth is expected to occur alongside galaxy growth as galaxies quench. Galaxies can evolve rapidly from star-forming to quiescent while changing morphology if they undergo a gas-rich major merger. These mergers will trigger starbursts as well as AGN activity, providing a means for AGN feedback to assist in quenching and preserving the black hole -- bulge relations \citep{Hopkins2008}. However, the starburst and peak of black hole growth cannot be simultaneous, due to the angular momentum loss required for the gas fueling star formation to reach the nucleus. The peak in AGN activity may be 100 Myr to more than a Gyr after the peak in starburst activity \citep{Davies2007, Schawinski2009, Wild2010, Hopkins2012b, Cales2015, Ellison2025}. If AGN activity acts to preserve the black hole -- bulge relation, this predicts that galaxies will experience a temporary offset from the black hole -- bulge relation as black hole growth lags behind the growth of the bulge during the starburst. 

This offset is difficult to observe due to the short timescale in which galaxies can be seen in a ``post-starburst" phase \citep{Wild2010, Snyder2011}. Post-starburst galaxies are typically bulge-dominated \citep{Yang2008} and have formed $\sim10$\% of their stellar mass in a recent starburst, indicating that their bulge components have also grown by $\sim10$\% \citep{French2018}. If AGN feedback acts to preserve the black hole -- bulge relation, we should observe young, gas-rich post-starbursts with under-massive black holes, and old, gas-poor post-starbursts to lie back on the black hole -- bulge relation. 

\begin{figure*}[t!]
\begin{center}
\includegraphics[width=1\textwidth]{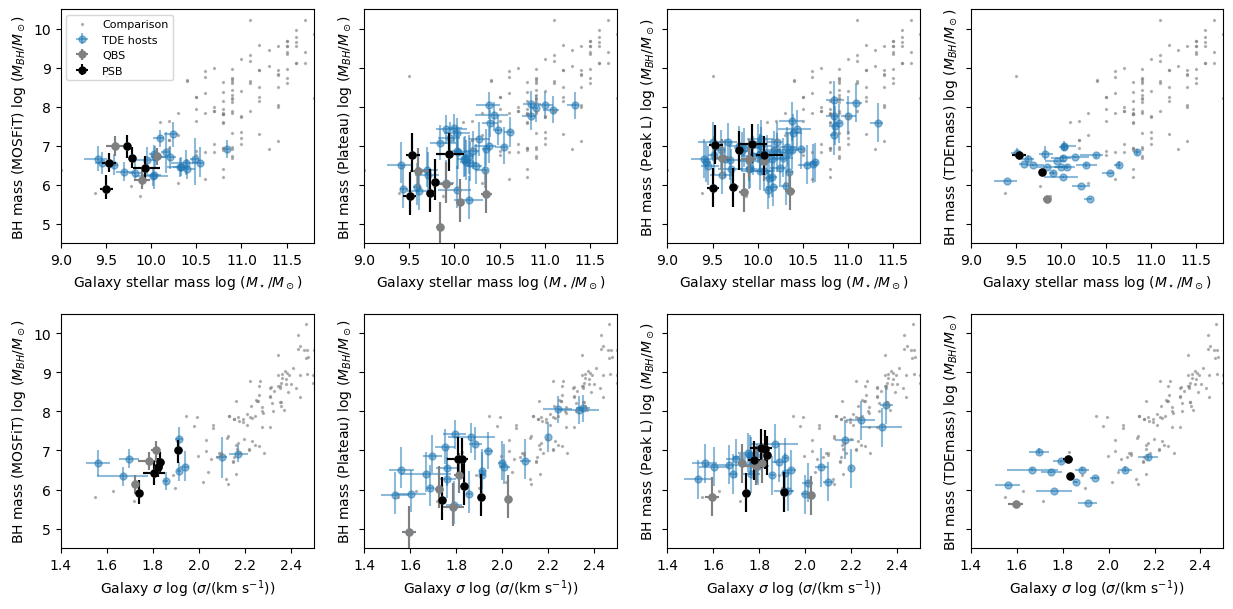}
\end{center}
\vspace{-5mm}
\caption{Comparison of host galaxy masses (top row) and bulge velocity dispersions (bottom row) vs. black hole masses inferred using a range of TDE techniques (blue). The left column shows black hole masses measured using \mosfit by \citet{Mockler2019, Nicholl2022}. The second column shows black hole masses measured using plateau luminosities, assuming this luminosity is driven by disk emission \citep{Mummery2024} and the third column shows black hole masses inferred from scaling relations between the plateau luminosity and peak luminosity \citep{Mummery2024}. The fourth column shows black hole masses measured by \citet{Hammerstein2023} using TDEmass \citep{Ryu2020}. Comparison galaxies from \citet{Greene2020} are shown as small grey squares. 
Here, we have labeled post-starburst galaxies (PSB) and a similar selection of quiescent Balmer-strong galaxies (QBS). These galaxies have high TDE rates and are strongly over-represented within the sample. The relative delay of AGN feedback with respect to quenching will affect whether PSB and QBS galaxies are offset on this relation. 
}
\label{fig:bh_bulge}
\end{figure*}

TDEs are an ideal way to measure black hole masses in these galaxies, as they are typically too far away for dynamical black hole mass measurements, and because post-starburst galaxies have a high TDE rate \citep{Arcavi2014, French2016, French2020}. In Figure \ref{fig:bh_bulge} we show the current distribution of post-starburst TDE hosts (and quiescent Balmer-strong TDE hosts, a slightly weaker post-starburst selection) on the black hole mass -- stellar mass and black hole -- bulge relations. Galaxy properties and TDE-inferred black hole masses have been compiled from \citet{Mockler2019, Nicholl2022, Hammerstein2023, Mummery2024} alongside comparison galaxies compiled by \citet{Greene2020}. Larger numbers will be required to assess any offsets, but we see tentative evidence for an offset in the plateau BH mass - velocity dispersion relation. Understanding the root cause for the post-starburst TDE rate enhancement will be important for this measurement, as TDEs from a secondary black hole  \citep{Cen2020, Melchor2024} could also appear as undermassive relative to their host galaxies. 

Over the first few years of Rubin, we expect to find thousands of TDEs, of which $\sim10-33$\% \citep{French2020} are expected to be in post-starburst galaxy hosts, yielding hundreds of black hole masses and bulge masses measured from Rubin data. Given the black hole mass measurement errors we predict here, we will not be able to search for this offset using individual galaxies, but will be able to see a net offset considering population averages.

\subsubsection{Using Event Horizon Suppression to Measure Black Hole Spins}

\begin{figure}
\begin{center}
\includegraphics[width=0.49 \textwidth]{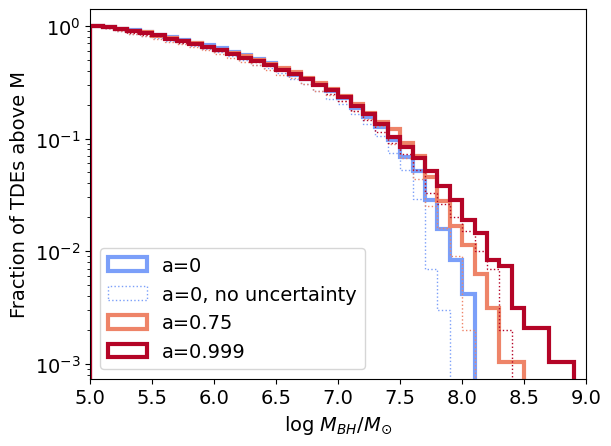}
\end{center}
\caption{Cumulative fraction (inverse) of TDEs observed for black holes with (solid lines) and without (dotted lines) measurement uncertainty of 0.26 dex, for a simulated sample of 1000 TDEs. We use the TDE rates as a function of black hole mass and spin from \citet{Mummery2024spin}, and show the results for three spin values. In both the cases with and without added uncertainty, the trend for higher spins to result in a larger number of TDEs with higher black hole mass can be seen. When we consider measurement uncertainty, Eddington bias results in a larger number of TDEs with high measured black hole masses, as there are more lower-mass systems that can be erroneously measured at higher black hole masses than the inverse. Even with 0.26 dex measurement errors, it is clear that the differing spins will have a measurable impact on the observed black hole masses.}
\label{fig:spin}
\end{figure}

The spin of supermassive black holes is difficult to measure, but is a key parameter both in setting the physics of the event horizon and tracing the growth mechanisms of the black hole. Coherent gas accretion through a long-lived thin disk will spin up black holes, while chaotic accretion \citep{King2006} can act to decrease the spin; mergers will often spin up low $a$ black holes and spin down high $a$ black holes \citep[e.g.,][]{Volonteri2005, Griffin2019}. The dominant mode of accretion and the role of mergers relates sensitively to the host galaxy. If high mass black holes have grown primarily through long-lived episodes of coherent accretion, and lower mass black holes are more easily affected by mergers and new gas flows, we might expect higher mass black holes to have uniformly high spins while lower mass black holes have a range of more moderate spins \citep{Volonteri2007}. However, several recent works using simulations and semi-analytic models have proposed that chaotic accretion dominates at the high mass end and in dispersion-dominated galaxies (leading to lower spins), while coherent accretion (and higher spins) are seen in lower mass, rotation-dominated galaxies \citep{Sesana2014, Fiacconi2018, Bustamante2019}. In addition to tracing growth mechanisms, the spin of a black hole can affect the launching of jets and AGN feedback \citep{Fiacconi2018, Bustamante2019}.

However, supermassive black hole spins are difficult to measure, with only a few dozen constraints in the literature. Most supermassive black hole spins are high, consistent with a maximal spin of $a\sim1$; in a review of 32 supermassive black hole spin measurements compiled by \citet{Reynolds2021}, only 8 objects have spin upper limits $a<1$. However, the majority of these measurements are from the X-ray reflection method, which requires a geometrically thin, optically thick accretion disk and thus a relatively narrow range of Eddington ratios. Furthermore, the luminosity of the accretion disk will be higher at a given mass accretion rate for more rapidly-spinning black holes, and thus X-ray flux-limited selections will be biased towards more rapidly-spinning black holes \citep{Brenneman2011, Vasudevan2016}.

TDEs provide a new method for measuring supermassive black hole spins for a complimentary selection of supermassive black holes. For a TDE to be observable, the tidal radius must be outside the event horizon to produce an observable flare. For a main sequence star around a Schwarzschild black hole, this means TDEs will not be observed for black holes with mass $M_{BH}>10^8 \ M_\odot$. However, the spin of the black hole will affect the cutoff mass, with TDEs observable from black holes $M_{BH}>10^8-10^9 \ M_\odot$ with high spins ($a>0.9$) \citep{Kesden2012, Leloudas2016, Stone2019, Mummery2024spin}. Thus, the rates of TDEs in massive black hole systems can be used to infer the typical spin values in these systems. 

To model the impact of black hole mass measurement uncertainty on spin inference, we use the models from \citet{Mummery2024spin} for the estimated TDE rate as a function of black hole mass and spin. In Figure \ref{fig:spin}, we demonstrate the impact of measurement uncertainty on the cumulative fraction of TDEs seen for black holes above a given mass threshold. Drawing from the TDE rate functions of \citet{Mummery2024spin}, we model a population of 1000 TDEs, and the distribution of measured masses for a range of spin values. In general, $\lesssim 1$\% of TDEs will be from black hole masses $>10^8$ \Msun for any spin case, so a large number of TDEs (of order 1000) will be needed to search for differences from differing spins. We model the impact of measurement uncertainty by adding a Gaussian-distributed random value with $\sigma=0.26$ dex to match our uncertainty estimates above. In both the cases with and without added uncertainty, the trend for higher spins to result in a larger number of TDEs with higher black hole mass can be seen. When we consider measurement uncertainty, Eddington bias results in a larger number of TDEs with high measured black hole masses for all spins, as there are more lower-mass systems that can be erroneously measured at higher black hole masses than the inverse. Even with 0.26 dex measurement errors, it is clear that the differing spins will have a measurable impact. A full accounting of uncertainties will be needed to disentangle trends due to the underlying spin distribution from Eddington bias.

\section{Conclusions}
\label{sec:conclusions}

We use \mosfit and {\tt rubin-sim} to generate synthetic TDEs from a realistic distribution of black hole masses, star masses, and impact parameters, as they will be observed by Rubin. We aim to understand how the observational cadence and photometric uncertainties will combine with the inherent degeneracies in this model to produce uncertainty in the recovered parameters. 

\begin{itemize}
    \item Black hole masses can be recovered with typical errors of 0.26 dex. Constraints on the early shape of the lightcurve are useful in improving the black hole mass recovery.  If there are limits of at least 1 magnitude below the first detection in the 30 days prior to it, the typical black hole mass recovery improves slightly to 0.25 dex and extreme outliers $>1$ dex from the input mass are avoided.
    \item Recovery of the mass of the disrupted star is difficult, limited by the degeneracy with the accretion efficiency. The median difference between the input and recovered star mass is 0.24 dex.
    \item The impact parameter $b$ can be recovered with accuracy of 0.42 (0.15 dex). The difficulty in constraining the power-law slope of the decline of the bolometric luminosity likely contributes this this uncertainty, and thus UV photometry will help to provide more accurate $b$ measurements for Rubin TDEs. 57\% of the cases have accurate recovery of whether the events are full or partial, so we caution the use of best-fit $b$ measurements in Rubin observations to assess whether TDEs are partially or fully disrupted systems. 
\end{itemize}

Black hole mass measurements obtained from Rubin observations of TDEs will provide powerful constraints on the black hole mass function, black hole -- galaxy co-evolution, and the population of black hole spins, though continued work to understand the origin of TDE observables and how the TDE rate varies among galaxies will be necessarily to fully utilize the upcoming rich data set from Rubin. 

\begin{acknowledgments}
This research was supported in part by grant NSF PHY-2309135 to the Kavli Institute for Theoretical Physics (KITP). We thank the referee for their thoughtful comments, which have improved the clarity of this manuscript.
\end{acknowledgments}

\bibliography{refs}{}
\bibliographystyle{aasjournal}

\end{document}